# The role of resonances in the capture of $^8$Li$(p,\gamma)^9$Be on the reaction rate of the relevant astrophysical synthesis of $^9$Be

## Dubovichenko S.B.$^{1,2,*}$, Burkova N.A.$^2$ and Dzhazairov-Kakhramanov A.V.$^{1,*}$

$^1$ *Fesenkov Astrophysical Institute "NCSRT" ASA MDASI RK, 050020, Almaty, Kazakhstan*
$^2$ *al-Farabi Kazakh National University, RK, 050040, Almaty, Kazakhstan*

**Abstract:**

The total cross sections of the radiative proton capture on $^8$Li at astrophysical energies are considered in the framework of the modified potential cluster model with forbidden states, with the classification of the orbital cluster states according to Young diagrams. The data obtained from the $^9$Be$(\gamma,p_0)^8$Li photodisintegration process according to the detailed balancing principle were used for the experimental data of the $^8$Li$(p,\gamma)^9$Be capture. Overall, it is possible to obtain the available data on cross sections at energies to 7.0 MeV. Astrophysical $S$-factors and reaction rates $\langle\sigma v\rangle$ at the temperature range of 0.01 to 10 $T_9$ are calculated. It is shown that the resonances in the $p^8$Li scattering channel considerably influence the reaction rate, beginning with the first resonance at 87 keV. The analytical parametrization is obtained for the calculated reaction rate $\langle\sigma v\rangle$.

## 1. Introduction

The study of the formation mechanisms of $^9$Be directly concerns the problem of the overlap of the $A = 8$ mass gap and the synthesis of heavier elements in the early Universe, as well as the $r$-process nucleosynthesis in supernovae (see, for example, [1,2]). In the present time, it is considered that $^9$Be is formed as a result of a two-stage process: the radiative capture of alpha particles $\alpha(\alpha,\gamma)^8$Be, leading to the synthesis of the short-half-life isotope $^8$Be ($t_{1/2} = 6.7 \times 10^{-17}$s), and radiative neutron capture $^8$Be$(n,\gamma)^9$Be [2,3]. In addition, there is the more complicated process of the direct three-particle capture $\alpha\alpha n \to \gamma^9$Be, (see, for example, [4–8]).

Simultaneously, in [8], different binary channels of $^9$Be photodisintegration, namely, $^9$Be$(\gamma,p)^8$Li, $^9$Be$(\gamma,d)^7$Li, $^9$Be$(\gamma,t)^6$Li and also $^9$Be$(\gamma,^3$He$)^6$He, were experimentally studied. It is evident that the processes of two-body radiative capture related with them by the detailed balancing principle lead to the synthesis of $^9$Be and require the corresponding estimation contextually in the astrophysical supplements. Meanwhile, it should be noted that the first three reactions have the Coulomb barrier in channels $p^8$Li, $d^7$Li and $t^6$Li lower than in $^3$He$^6$He along with $^4$He$^5$He channel, namely, in the ratio of 3:4. Thus, proton, deuteron and triton ($Z = 1$) radiative capture induced reactions on lithium isotopes ($Z = 3$) have a preference in charge combinations for the Coulomb burrier penetrability. In case of alpha-particle induced reactions a higher factor of the Coulomb suppression takes place in the initial channels of the $\alpha(\alpha,\gamma)^8$Be and $\alpha\alpha n \to \gamma^9$Be processes.

---

$^*$Corresponding authors: albert-j@yandex.ru (A.V. Dzhazairov-Kakhramanov),
dubovichenko@mail.ru (S.B. Dubovichenko).

Here, we consider the reaction $^8$Li$(p,\gamma)^9$Be at low astrophysical energies in connection with those included in the chain of significant processes, which lead to the synthesis of heavier elements in the fundamentally significant work of Terasawa et al. [1], which has played the role of certain "proposal practical research guide" from the moment of its publishing almost 20 years ago.

The cross section of $^8$Li$(p,\gamma)^9$Be is hard to obtain directly due to low $^8$Li beam intensity and the small cross section at astrophysical energies. In addition, the difficulty of studying the reaction of $^8$Li$(p,\gamma)^9$Be also lies in the fact that the direct experimental measurement of cross sections is not easy task due to the very short half-life of $^8$Li – 838 ms [9]. However, as in the case of the neutron capture on $^8$Li [10], some indirect methods of extracting the direct capture cross sections can be used with the help of the radiative capture model and spectroscopic factor frame [9,11].

The $p^8$Li$\to^9$Be$\gamma$ reaction presents significant astrophysical interest because it is included in the chains of primordial nucleosynthesis of the early Universe [12]. However, its experimental investigation in the astrophysical energy range has so far been insufficient. Apparently, there is only one work [13] where this range is considered. It is also supplemented, by data from [14], where measurements were carried out at higher energies. However, these works were published in the 1960s and we do not currently possess more modern experimental studies of the total cross sections of the considered reaction [15]. This is in spite of the fact that studies of spectra $^9$Be in the $p^8$Li channel are still continuing [16]. It should be noted, the available and considered further theoretical results differ so greatly that it is difficult to draw certain conclusions regarding the rate of this reaction. Meanwhile, all these early calculations do not take into account the existence of the resonances in the $p^8$Li scattering states at low energies [16].

In the present study, we consider the reaction of the proton capture on $^8$Li in the frame of the modified potential cluster model (MPCM) [17,18] and define how the criteria of this model allow us to correctly describe the total cross sections and the astrophysical $S$-factor at astrophysical energies. The energy range of 10 keV to 7.0 MeV is considered but only taking into account the structure of resonances up to 2.0 MeV, as discussed previously [16]. The reaction rate is calculated at the temperature range of 0.01 to 10 $T_9$. The analysis of the influence of the location and resonance widths onto the value and shape of the reaction rate is presented.

## 2. Model and calculation methods

Further we use well-known formulas for the total cross sections and partial matrix elements of $NJ$ transition operators [17–19] ($S_i = S_f = S$ – total spins)

$$\sigma_c(NJ, J_f) = \frac{8\pi Ke^2}{\hbar^2 k^3} \frac{\mu \cdot A_J^2(NJ,K)}{(2S_1+1)(2S_2+1)} \frac{J+1}{J[(2J+1)!!]^2} \sum_{L_i,J_i} P_J^2(NJ,J_f,J_i) I_J^2(J_f,J_i), \quad (1)$$

here $K = E_\gamma/(\hbar c)$ is the wave number of the emitted photon with energy $E_\gamma$, $\mu$ is the reduced mass, $k$ is wave number of particles in the initial channel, $e$ is the elementary charge.

Matrix elements of the electric $EJ$ transitions have a form

$$P_J^2(EJ, J_f, J_i) = \delta_{S_iS_f}[(2J+1)(2L_i+1)(2J_i+1)(2J_f+1)](L_i 0 J 0 | L_f 0)^2 \begin{Bmatrix} L_i & S & J_i \\ J_f & J & L_f \end{Bmatrix}^2$$

$$A_J(EJ, K) = K^J \mu^J \left( \frac{Z_1}{m_1^J} + (-1)^J \frac{Z_2}{m_2^J} \right), \qquad I_J(J_f, J_i) = \langle \chi_f | r^J | \chi_i \rangle \qquad (2)$$

Here $L_f$, $L_i$, $J_f$, $J_i$ – are momenta of initial ($i$) and final ($f$) channel particles, $Z_1$, $Z_2$, $m_1$, $m_2$ are charges and masses of the initial channel particles. The integral $I_J(J_f, J_i)$ is defined by using wave functions of relative motion of particles in the initial $\chi_i$ and final $\chi_f$ states, which depend on an intercluster distance $r$.

Matrix elements of the magnetic process $M1$ at $J = 1$ are written as ($S_i = S_f = S$, $L_i = L_f = L$)

$$P_1^2(M1, J_f, J_i) = \delta_{S_iS_f}\delta_{L_iL_f}[S(S+1)(2S+1)(2J_i+1)(2J_f+1)] \begin{Bmatrix} S & L & J_i \\ J_f & 1 & S \end{Bmatrix}^2,$$

$$A_1(M1, K) = \frac{\hbar K}{m_0 c} \sqrt{3} \left[ \mu_1 \frac{m_2}{m} - \mu_2 \frac{m_1}{m} \right], \qquad I_J(J_f, J_i) = \langle \chi_f | r^{J-1} | \chi_i \rangle. \qquad (3)$$

Here, $m$ is the mass of the synthesized nucleus, $\mu_1$, $\mu_2$ are the magnetic momenta of the clusters, and the remaining notations are given as in the previous expression.

Constant $\hbar^2/m_0$ is equal to 41.4686 MeV fm$^2$, where $m_0$ is the atomic mass unit (amu). The point-like Coulomb potential is written in the form $V_{Coul} = 1.439975\, Z_1Z_2/r$, where $r$ is the relative distance between particles of the initial channel in fm. The magnetic momenta of the proton $\mu_p = 2.792847\mu_0$ [35] and $^8$Li nucleus $\mu(^8\text{Li}) = 1.65356\mu_0$ [20], where $\mu_0$ is the nuclear magneton.

## 3. Classification of $p^8$Li states according to Young diagrams

For many cluster systems, we are basing on the concept of Young diagrams {*f*} classification, see for details in [17,18] and, for example, in reviews [22,26]. As it was shown in [21] the Young diagram {431} is corresponded to the ground state (GS) of $^8$Li in the $n^7$Li channel.

Therefore, for the $p^8$Li system, we have the following set of Young diagrams as the result of direct product: {431} + {1} = {531} + {441} + {432} + {4311}. The first diagram {531} is forbidden because it cannot be five nucleons in the *s*-shell – it corresponds to the orbital momenta $L = 1, 2, 3$, determined by the Elliot rule [23]. The second diagram {441} corresponds to $L = 1, 2, 3, 4$, the third one {432} has $L = 1, 2, 3$ and the fourth {4311} corresponds to $L = 0, 2$. The second diagram {441} we will consider allowed, and put it in the correspondence to the GS of $^9$Be in the $p^8$Li channel [24]. Diagrams {432} and {4311} are not considered because without product tables of Young diagrams [25], it is impossible to understand if they are forbidden or allowed.

From the given classification it follows that for the $p^8$Li system (it is known that $J^\pi, T = 2^+, 1$ for $^8$Li [24]) in potentials of $S$ waves there is no forbidden state (FS). In the case of *P* waves for diagram {441}, there are both FS and allowed state (AS). In the *D* and *F* states there is FS for the same diagram, which can be considered as bound. The state in the $P_{3/2}$ wave corresponds to the GS of $^9$Be with $J^\pi, T = 3/2^-, 3/2$ and lays at the binding energy of $p^8$Li system of $E_b = -16.8882$ MeV [24]. Some of $p^8$Li scattering states and bound states (BSs) can be mixed by spin channel with $S = 3/2$ and $5/2$. As far $p^8$Li and $n^8$Li are isobar analogue systems, it is reasonable to consider them in parallel basing on the previous experience of the detailed analysis of $n^8$Li in [10]. As a consequence, we assume that the GS of $^9$Be in the $p^8$Li channel corresponds to the $^4P_{3/2}$ state (in spectroscopic notations $^{2S+1}L_J$).

Once, we are following the parallel analysis of $n^8$Li and $p^8$Li systems, it should be noted that earlier in [10], we considered another option of classification $n^8$Li system for $^9$Li. In particular, the Young diagram {44} was used for $^8$Li, then for the $n^8$Li system, we obtain {1} + {44} = {54} + {441} [23]. The first diagram {54}, forbidden according to the Elliot rule, corresponds to the orbital momenta $L = 0, 2, 4$, and the second one {441} is allowed with $L = 1, 2, 3, 4$ [23]. In this case, there is a FS in the *S* and *D* waves and only AS in the *P* wave. Such ambiguity for Young diagrams of $A = 8$ takes place due to the fact that in the channel $N + {}^7A$ with the channel spin $S = 2$ in the frame of 1*p*-shell there is no allowed Young diagram at all [26,27].

## 4. Structure of the $p^8$Li resonance states

Let us discuss now the state structure of $^9$Be, where we consider further the GS and six low-lying resonance states. The GS of $^9$Be at the $E_b$ energy (round off energy to 1 keV) [16] relatively to the threshold of the $p^8$Li channel and is matched to the $^4P_{3/2}$ state. So that, the *E*1 capture is possible from $^4S_{3/2}$ scattering wave to the $^4P_{3/2}$ GS of $^9$Be, that is, the dominant contribution is given due to the transition $^4S_{3/2} \to {}^4P_{3/2}$.

The spectrum of the resonance states of $^9$Be in the cluster $p^8$Li channel is listed in Table 1.

Table 1. Comparison of data on the $^9$Be spectrum from different works. The threshold of the $p^8$Li channel in $^9$Be equals $E_b$ [24]. The results coinciding for all three works are marked by bold.

| [16] 2018 year | | | [28] 2016 year | | | [24] 2004 year | | |
|---|---|---|---|---|---|---|---|---|
| $E_r$, MeV | $J^\pi$ | $\Gamma_{c.m.}$, keV | $E_r$, MeV | $J^\pi$ | $\Gamma_{c.m.}$, keV | $E_r$, MeV | $J^\pi$ | $\Gamma_{c.m.}$, keV |
| – | – | – | 0.087(1) | 1/2$^-$ | 0.39(1) | 0.0868(8) | 1/2$^-$ | 0.389(1) |
| **0.420(7)** | **5/2$^-$** | **210(20)** | **0.410(7)** | **5/2$^-$** | **195** | **0.410(7)** | **5/2$^-$** | **200** |
| **0.610(7)** | **7/2$^+$** | **47(7)** | **0.605(7)** | **7/2$^+$** | **47** | **0.605(7)** | **7/2$^+$** | **47** |
| 1.100(30) | 3/2$^+$ | 50(22) | 1.132(50) | – | – | 1.132(50) | – | – |
| 1.650(40) | 7/2$^-$ | 495(34) | 1.688(30) | 5/2$^+$ | 432(50) | 1.692(40) | – | – |
| 1.800(40) | 5/2$^-$ | 79(17) | 1.758(40) | 7/2$^+$ | 490(81) | 1.762(50) | 5/2$^-$ | 300(100) |
| – | – | – | 2.352(50) | 3/2$^+$ | 310(80) | 2.312(50) | – | 310(80) |

It is reasonable to discuss the spectrum data more thoroughly as there are some discrepancies coming from the different published sources. Our standpoint is that the systematization, given below, allows *a priori* to estimate the most significant contribution of resonance states in the capture cross section (the resonance states included in consideration are marked by bold italics).

   1. The ***first resonance state*** (**1$^{st}$ RS**) is located at 16.975(8) MeV relative to the GS or 0.0868(8) MeV [24] (0.087(1) MeV [28]) in the center of mass (c.m.) relative to the threshold of the $p^8$Li channel. $J^\pi = 1/2^-$ is given for this level [24,28] that allows us to take $L = 1$ for it, that is, to consider it quartet $^4P_{1/2}$ resonance. The level width of $\Gamma_{c.m.} = 0.39(1)$ keV is given in [24,28]. It is possible to construct the unambiguous $^4P_{1/2}$ potential of the elastic scattering according these data. The *M*1 transition $^4P_{1/2} \to {}^4P_{3/2}$ to the GS is allowed for this state.

   2. The ***second resonance state*** (**2$^{nd}$ RS**) is located at 17.298(7) MeV relative to the GS or 0.410(7) MeV in the c.m. relative to the threshold of the $p^8$Li channel [24,28]. $J^\pi = 5/2^-$ is given for this level [24,28] that allows us to take $L = 1$ for it, that is, to consider it quartet $^4P_{5/2}$ resonance. The level width of $\Gamma_{c.m.} = 200$ keV is given in [24] and 195 keV in [28]. The energy 420(7) keV at the width 210(20) keV is given in new work [16]. The *M*1 transition $^4P_{5/2} \to {}^4P_{3/2}$ to the GS is also allowed for this state.

   3. The *third resonance state* (3$^{rd}$ RS) is located at 17.493(7) relative to the GS or 0.605(7) MeV in the c.m. relative to the threshold of the $p^8$Li channel with the width of 47 keV [24,28]. $J^\pi = 7/2^+$ is given for this level [24,28] that allows one to consider it as a quartet $^4D_{7/2}$ resonance, but in this case only the *E*2 transition to the GS of $^9$Be is allowed, which we will not consider due to its small contribution.

   4. The ***fourth resonance state*** (**4$^{th}$ RS**) at energy 1.100 MeV with the width 50(22) keV and momentum 3/2$^+$, which is in data [16], can be considered as refinement of data from [24,28]. Neither its width nor its momentum is not given in them, only the energy of 1.132(50) MeV that approximately coincides with new results from [16] is given. If to take the $^4D_{3/2}$ state for it, the *E*1 transition $^4D_{3/2} \to {}^4P_{3/2}$ to the GS turn out to be possible. We failed to obtain the resonance in the $^4S_{3/2}$ wave with such characteristics; therefore, we consider this wave nonresonance.

   5. The *fifth resonance state* (5$^{th}$ RS) according to data [16], the energy is equal to 1.650(40) MeV and the width of 495(34) keV that coincide with data from [28], but the value 7/2$^-$ is given for momentum. If to assume that the last momentum corresponds to 5$^{th}$ RS, so it can refer to the $^4F_{7/2}$ scattering state, and then only *E*2 transition to the GS is allowed. It will not be considered.

   6. The *sixth resonance state* (6$^{th}$ RS) according to [24] is located at the excitation energy of 18.65(5) MeV or 1.762(50) MeV in the c.m. relatively to the threshold of the $p^8$Li channel with the width 300(100) keV. $J^\pi = 5/2^-$ is given for this level [24] that allows us to take $L = 1$ for it, i.e., to consider it quartet $^4P_{5/2}$ resonance. In [16], the similar energy of 1.800(40) MeV with slightly smaller width of 79(17) keV and momentum 5/2$^-$ are given. If to recognize this momentum, it can be referred to the $^4P_{5/2}$ scattering state and then the *M*1 transition $^4P_{5/2} \to {}^4P_{3/2}$ to the GS is allowed. However, we do not succeed to obtain characteristics of the *P* wave of continuous spectrum that is noted in Table 1.

   7. The *seventh resonance state* (7$^{th}$ RS) with the energy of 2.312(50) with the width 310(80) keV and unknown momentum are given in [24]. The energy of 2.352(50) with the same width and momentum 3/2$^+$ is given in [28]. As we have seen above, the results of [28] on two previous levels differ from new data [16]; therefore, the data for this level most probably should be specified, and now we will not consider this resonance.

Slightly higher at excitation energy of 19.420(50) MeV, there is other resonance with the width of 600(100) keV, but it has a presumable momentum $9/2^+$ [28] and cannot lead to $E1$ or $M1$ transitions and we therefore do not consider it. Furthermore, we will base on spectra given in new work [16], including first resonance at 87 keV from [24,28] and limiting by the energies not higher the threshold of 2 MeV.

Thus, the basic transitions to the GS that are considered here and also $P^2$ coefficients for the total cross sections from relations (2)–(3), are presented in Table 2. The known values of isotopic spin $2T_i$ for the resonance states with $J_i^\pi$ are given.

Table 2. Characteristics of taken into account transitions at the $^8$Li(p,γ)$^9$Be capture.

| No. | $[^{2S+1}L_J]_i$ | Resonance energy, MeV | $J_i^\pi$, $2T_i$ | Transition | $[^{2S+1}L_J]_f$, $2T_f = 1$ | $P^2$ |
|---|---|---|---|---|---|---|
| 1. | $^4S_{3/2}$ | – | $3/2^+$ (?) | $E1$ | $^4P_{3/2}$ | 4 |
| 2. | $^4D_{3/2}$ | 1.100 (4$^{th}$ RS) | $3/2^+$, (3?) | $E1$ | $^4P_{3/2}$ | 64/25 |
| 3. | $^4P_{1/2}$ | 0.087 (1$^{st}$ RS) | $1/2^-$, 3 | $M1$ | $^4P_{3/2}$ | 10/3 |
| 4. | $^4P_{5/2}$ | 0.410 (2$^{nd}$ RS) | $5/2^-$, 3 | $M1$ | $^4P_{3/2}$ | 18/5 |

Let us discuss other important aspect of the approaches used by us. As seen in Table 2, for the **1$^{st}$ RS** (0.087) and the **2$^{nd}$ RS** (0.410), the isospin $T = 3/2$. This means that for these states there are no channel mixing $p + ^8$Li, $d + ^7$Li, and also $\alpha + ^5$He, which have the isospin $T = 1/2$. Problems of channels coupling are discussed in [29,30]. Thus, the single-channel approach, which we use here, is quite justified that capture from **1$^{st}$ RS** (0.087) is the strongest one (see further).

## 5. Criteria of the potential construction

We now discuss the concepts of construction of interaction potentials in the frame of the MPCM. Although we did not succeed to find the results of the phase shift analysis of the $p^8$Li elastic scattering and they are unlikely to exist, potentials are constructed on the basis of $^9$Be spectra. Thus, $S$ potentials, since in $^9$Be spectra at considering energies there are no $^4S$ resonances, should lead to the $^4S$ scattering phase shifts on the point of zero or slowly decreasing $S$ phase shifts. These are relatively weak criteria for the construction of the potential, therefore its parameters can be considered as variable in the present model. The potentials of resonance waves are constructed in order to correctly describe the location of resonance $E_r$ and its width $\Gamma_{c.m.}$, therefore their parameters are obtained quite unambiguously.

GS potentials are constructed in such a form that allows one to correctly describe the channel binding energy, the charge radius of $^9$Be and its asymptotic constant in the $p^8$Li channel. Since, all known values of the asymptotic normalization coefficient ($A_{NC}$) and the spectroscopic factor $S_f$, according to which the asymptotic constant (AC) is obtained, have enough large error. The GS potentials also can have few options with different parameters of width. However, at the given values of AC and binding energy, its parameters are obtained completely unambiguously. The procedure of potential construction was described in [19] to some detail.

The radius of $^8$Li equals 2.327±0.0298 fm, which is given in the database [31], is used in present calculations. The radius of 2.518±0.0119 fm for $^9$Be also is known from the database [31]. In addition, for example, the value of 2.299(32) fm for $^8$Li radius was found in [32]. In [33], for these radii, values of 2.30(4) and 2.519(12) fm were obtained, correspondingly. All these data agree well among themselves within the limits of errors. The exact values of $m(^8Li) = 8.022487$ amu [34] and $m_p = 1.0072764669$ amu [35] were used for the masses of $^8$Li nucleus and proton. The charge and mass radius of the proton is equal to 0.8775(51) fm [35]. For 1 amu energy equivalent of 931.4941024 MeV is used [35].

The spectroscopic factor $S_f$ of the GS and $A_{NC}$ are related as [36]:

$$A_{NC}^2 = S_f \times C^2, \tag{4}$$

where $C$ is the dimensioned asymptotic constant in fm$^{-1/2}$. Constant $C$ is related with dimensionless $C_w$ by $C = \sqrt{2k_0} C_w$, and the dimensionless constant $C_w$ defined from the asymptotic equation [37]:

$$\chi_L(r) = \sqrt{2k_0} C_w W_{-\eta L+1/2}(2k_0 r). \tag{5}$$

Here $\chi_L(r)$ is the numerical BS radial wave function, obtained as the solution of the Schrödinger equation normalized to unit size; $W_{-\eta L+1/2}(2k_0 r)$ is the Whittaker function of the bound state, determining the asymptotic behavior of the wave function and obtained as the solution of the same equation without nuclear potential; $k_0$ is a wavenumber related to the channel binding energy $E$ where $k_0 = \sqrt{2\mu E/\hbar^2}$ in fm$^{-1}$; $\eta$ is the Coulomb parameter $\eta = \mu Z_1 Z_2 e^2/(\hbar^2 k_0) = 3.44476 \cdot 10^{-2} \mu Z_1 Z_2 / k_0$.

Note that the spectroscopic factor $S_f$ is used only for the standard procedure of the obtaining possible $C_w$ range from the extracted in the experiment $A_{NC}$ value [19,36,37].

## 6. Potentials of the $p^8$Li interaction

As in our previous works [17–19] for other nuclear systems, we use the central potential of the Gaussian form with the given orbital momentum $L$ in each partial wave as the $p^8$Li interaction:

$$V(r,L) = -V_L \exp(-\gamma_L r^2). \tag{6}$$

The $^4P_{3/2}$ level we consider as the ground state of $^9$Be in the $p^8$Li channel and such potential should correctly describe the AC for this system. In order to extract this constant $C$ in form (4) or $C_w$ (5) from the available experimental data, let us consider information on the spectroscopic factors $S_f$ and asymptotic normalization coefficients $A_{NC}$. In [38], except for their own results, the authors add data of previous works. The selected data on the spectroscopic factors of close values are presented in Table 3.

Table 3. Data on spectroscopic factors $S_f$ for the GS of $^9$Be in the $p^8$Li channel from works [9,11,38,39]. $\overline{S}_f$ is the average value on data interval.

| Reaction from what $S_f$ was determined | $S_f$ for the $p^8\text{Li}_{GS}$ channel | Ref. |
|---|---|---|
| $^8\text{Li}(d,n)^9\text{Be}$ | 0.64(21) | [38] |
| $^{12}\text{C}(^9\text{Be},^8\text{Li})^{14}\text{N}$ | 0.73(15) | [11] |
| Average value | 0.69, that is, $\sqrt{S_f} = 0.83$ | |
| Average value on data interval | $\overline{S}_f = 0.66(22), 0.43–0.88$ $\sqrt{\overline{S}_f} = 0.81(14)$ | |
| $^9\text{Be}(^8\text{Li},^9\text{Be})^8\text{Li}$ | 1.50(28) | [39] |
| Potential model | 1.50(27) | [9] |
| Average value on all results | 1.09, $\sqrt{S_f} = 1.05$ | |
| Data interval on all results | $\overline{S}_f = 1.11(68), 0.43-1.78$ $\sqrt{\overline{S}_f} = 1.05(28)$ | |

Furthermore, the $A_{NC}$ of the $^4P_{3/2}$ GS in the $p^8$Li channel was given in [40], where $A_{NC} = 10.75(12)$ fm$^{-1/2}$ was obtained. The constant for the $^6P_{3/2}$ state is much less – 0.25(10) fm$^{-1/2}$ [40]. We consider here only one spin channel with $S = 3/2$. On the basis of expression (4) and average value on interval $\sqrt{S_f} = 0.66(22)$ from Table 2 according [11,38] for the AC GS, the value $C = 13.71(2.52)$ fm$^{-1/2}$ was obtained, and because $\sqrt{2k_0} = 1.307$ fm$^{-1/2}$, so dimensionless AC (5) is equal to $C_w = 10.49(1.93)$. If to use the average value $\sqrt{S_f}$ from Table 3 equals 1.05(28), then for constant $C$, we obtain a wider data interval 11.06(3.06) fm$^{-1/2}$ and $C_w = 8.46(2.34)$. Consequently, the possible interval of $C_w$ values on two data groups from Table 3 is approximately from 6 (from the latest data) to 12.5 (from the previous results).

Furthermore, two options of the GS potentials with FS, which allow us to obtain the dimensionless asymptotic constant $C_w$ in the given above limits, were obtained. The parameters of these potentials $V_L$ and $\gamma_L$, and also main characteristics of $^9$Be nucleus, obtained with them (binding energy $E_b$, asymptotic constant $C_w$, mass radius $<R>_m$ and charge radius $<R>_{ch}$) are listed in Table 4.

Table 4. GS potential parameters and main characteristics of $^9$Be

| No. | $V_L$, MeV | $\gamma_L$, fm$^{-2}$ | $E_b$, MeV | $C_w$ | $<R>_m$, fm | $<R>_{ch}$, fm |
|---|---|---|---|---|---|---|
| 1. | 212.151135 | 0.17 | -16.8882 | 10.2(1) | 2.40 | 2.46 |
| 2. | 286.178045 | 0.25 | -16.8882 | 6.6(1) | 2.36 | 2.38 |

Potential No. 1 has the FS and leads to the binding energy -16.8882 MeV at the accuracy of the used here finite-difference method for finding an energy of 10$^{-5}$ MeV [41]. The definition of the calculation expressions used here for the radii is given, for example, in [17]. The above AC errors are defined by their averaging over the distance interval from 6–8 to 10–12 fm, which is the AC stabilization range. The phase shift of the elastic scattering of this potential for the GS $^4P_{3/2}$ smoothly decreases down to zero and at 5.0 MeV has the value of ~330°. Here, we consider that in the presence of two bound FS and AS, the phase shift according to generalized Levinson theorem starting from 360° [23]. Another option of the GS potential that leads to a smaller AC is given in Table 4 (set No. 2).

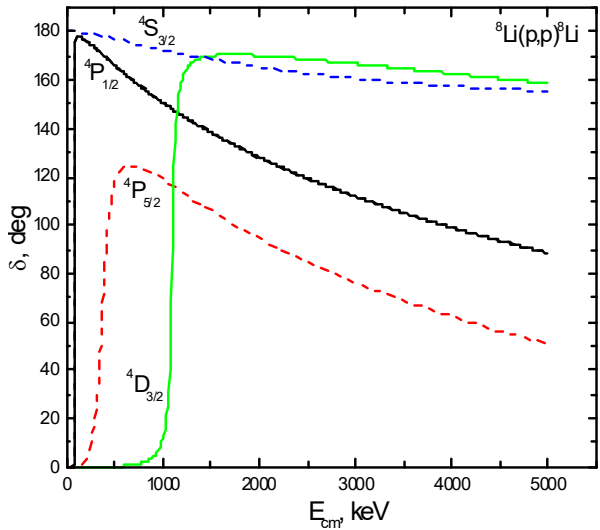

Fig. 1. Phase shifts of the elastic $p^8$Li scattering at low energies.

The parameters used by the scattering potentials are listed in Table 5. Potential No. 1 without FS was obtained for the S wave, which at energies up to 5.0 MeV leads to the scattering phase shift of ~155°, as shown by the blue dashed curve in Fig. 1. Here and further, it should be kept in view that in accordance with the generalized Levinson theorem [23], the phase shift of the potential without FS starting from 0°, but resonance potentials with one FS – from 180°. Fig. 1 shows the phase shifts without their renormalization according Levinson theorem.

Table 5. Options of potential parameters with FS for resonance states of nuclear and some characteristics obtained with them. In the two last columns, the experimental values listed above in Table 1 are shown.

| No. | $^{2s+1}L_J$ | $V_J$, MeV | $\gamma_J$, fm$^{-2}$ | $E_r$ (c.m.), keV present | $\Gamma_{c.m.}$, keV present | $E_r$ (c.m.), keV | $\Gamma_{c.m.}$, keV |
|---|---|---|---|---|---|---|---|
| 1. | $^4S_{3/2}$ | -5 | 0.1 | – | – | – | – |
| 2. | $^4D_{3/2}$ | 269.242 | 0.2 | 1100 | 56 | 1100(30) | 50(22) |
| 3. | $^4P_{1/2}$ | 66.69121 | 0.075 | 87 | ~0.4 | 87(1) | 0.39(1) |
| 4. | $^4P_{5/2}$ | 34.0399 | 0.04 | 410 | 203 | 420(7) | 210(20) |

Potential No. 3 from Table 5 leads to the $^4P_{1/2}$ scattering phase shift, plotted in Fig. 1 by the black solid curve and has the resonance at 87 keV. Scattering potential No. 4 corresponds to the phase shift $^4P_{5/2}$ presented in Fig. 1 by the red dashed curve and at the considered energies has the resonance at 410 keV. Potential No. 2 leads to the $^4D_{3/2}$ phase shift shown in Fig. 1 by the green solid curve. All resonance potentials have the bound FS, and their phase shifts are of 90.0°(1) at the resonance energy.

The $^4P_{3/2}$ phase shift is not given in Fig. 1. It was calculated with the GS potential parameters (Table 2). For both sets it smoothly decreases down and at 5.0 MeV has the value of ~330°. Here, we consider that in the presence of bound FS and AS the $^4P_{3/2}$ phase shift is starting from 360° according the generalized Levinson theorem [23].

## 7. Total cross sections, S-factor and reaction rate of radiative $^8$Li($p,\gamma$)$^9$Be capture

Let us discuss calculation results of the total cross sections for the proton capture on $^8$Li and comparing them with experimental measurements from [13,14]. The summed cross section is shown in Fig. 2a by the blue solid curve, black dashed curve shows the nonresonance capture from the $S$ wave with potential No. 1 from Table 5 to the GS with potential No. 1 from Table 4. The green dotted curve shows cross section at the capture from the first and the second resonances for potential No. 3 and No. 4 from Table 5. The red dashed curve shows the capture from the third resonance with potential No.2 from Table 5. The parameters of the $S$ scattering wave without FS were determined so as correctly reproduce total cross sections in the "plateau" range, which occurs at energies of 0.5 to 7 MeV. The potential No. 1 from Table 4 was used as the GS potential that reproduce AC from [40] to the best advantage.

It should be noted that there is no second resonance at 410 keV in the available data [13]. However, it exists in spectra [16,24,28], we are taking it into account. A new resonance at 410 keV not observed in experiment [13] is clearly observed in Fig. 2a. In this experiment, at this energy, the cross section minimum is observed, which is difficult to explain from the viewpoint of new data on $^9$Be spectra [16]. However, it is necessary to remember that these experimental data were obtained in sixties of the past century and were not confirmed later by more modern methods.

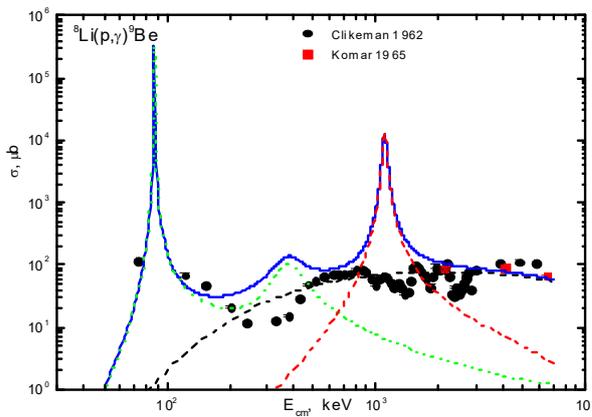
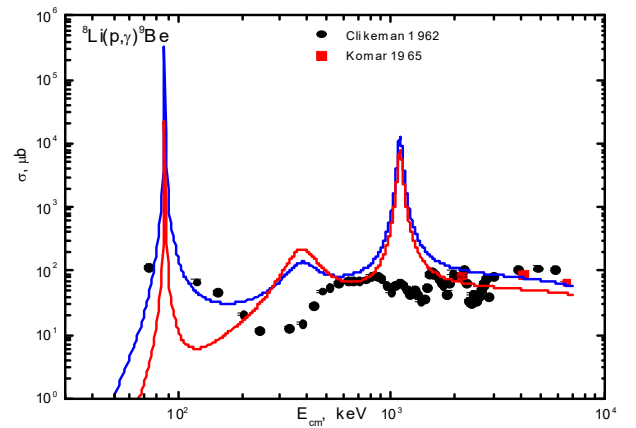

Fig. 2a Total cross sections of the radiative $^8$Li($p,\gamma$)$^9$Be capture to the GS of $^9$Be. Black points are experimental data from [13] and red squares are from [14]. Curves are explained in the text.

Fig. 2b Total cross sections of radiative $^8$Li($p,\gamma$)$^9$Be capture to the GS of $^9$Be. Notations like in Fig. 2a.

Now take note of the resonance at 1.1 MeV from Table 2, which can lead to the *E*1 transition to the GS. The calculation results that take into account this transition are shown in Fig. 2a. The sharp rise of the total cross section is observed at the resonance energy and is shown by the red dashed curve. This resonance also is not observed in the available experimental data.

The comparison of cross sections for different GS potentials with FS is shown in Fig. 2b. The previous results for potential No. 1 from Table 4 are shown by the blue curve and the results for potential No. 2 from Table 4 are shown by the red curve. These results are slightly lower than previous almost in the whole energy range and overall show less agreement with the available data [13].

Now, let us discuss calculation results of the astrophysical *S*-factor of the radiative capture for potentials from Tables 4 and 5. They are represented in Fig. 3 – the blue curve similar to Fig. 2a, and the green dotted curve with the value of 2.85(1) keV·b at 10 keV corresponds to the same curve in Fig. 2a, and black dashed curve corresponds to the same curve in Fig. 2a and shown nonresonance scattering from the *S* wave with the value of 1.69(1) keV·b at 10 keV. The calculated *S*-factor at 10 keV for both options of calculations is equal to 4.5(1) keV·b.

The value of 0.15 keV·b with a smooth decrease at 2.0 MeV down to 0.1 keV·b was obtained for the astrophysical *S*-factor in [38]. The interval of possible *S*-factor values at zero energy (*S*(0)) is determined to be 0.075 to 0.23 keV·b. The value of 0.3 keV·b for the *S*(0) was obtained in [39], and furthermore up to 1.5 MeV the *S*-factor smoothly decrease approximately down to 0.25 keV·b. The interval of possible *S*(0) defined in this work near the range of 0.21 to 0.38 keV·b. In [9] the *S*(0) is of 0.85 keV·b was obtained, and furthermore up to 1.5 MeV, the *S*-factor smoothly decreases approximately down to 0.75 keV·b. The interval of possible *S*(0) values defined near the range of 0.63 to 1.15 keV·b. However, in all of these works not any resonances, including the first one, have not been taken into account and the *S*-factor has a smooth shape. In addition, the results for comparison with experimental measurements [13,14] are not given in those works.

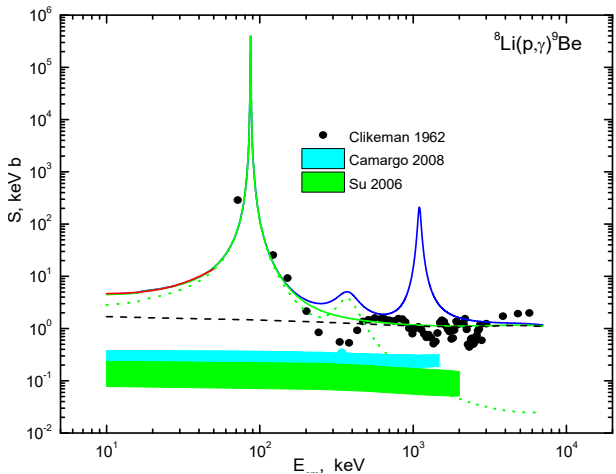

Fig. 3 Astrophysical *S*-factor of the radiative $^8$Li($p,\gamma$)$^9$Be capture to the GS of $^9$Be. Black points are experimental data from [13]. Curves are explained in the text.

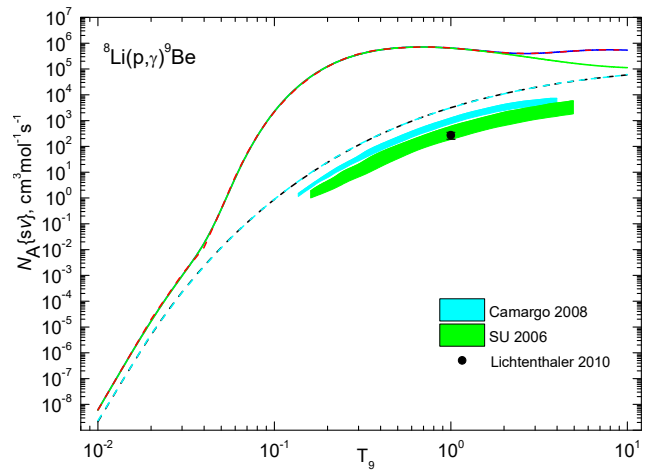

Fig. 4 Reaction rate of the proton capture on $^8$Li. Curves are explained in the text. Point is the result from [42].

The general parametrization of the *S*-factor in the energy range up to 50 keV was obtained by the expression of the form:

$$S(E_{c.m.}) = S_0 + S_1 E + S_2 E^2, \qquad (7)$$

where the energy is in keV and the corresponding parameters are $S_0 = 5.082$, $S_1 = -0.09391$ and $S_2 = 0.005359$. For these parameters, the $\chi^2$ value is equal to 0.26 with 5% errors comparing the initial calculated data. The result of this parametrization is shown in Fig. 3 by the red curve.

Furthermore, in Fig. 4, the green and blue solid curves show the reaction rate of the proton capture on $^8$Li, which corresponds to the same curves in Fig. 3 and is written in the form [43]:

$$N_A \langle \sigma v \rangle = 3.7313 \cdot 10^4 \mu^{-1/2} T_9^{-3/2} \int_0^\infty \sigma(E) E \exp(-11.605 E / T_9) dE, \qquad (8)$$

where $E$ is in MeV, the cross section $\sigma(E)$ is in μb, and $T_9$ is the temperature in $10^9$ K. Integration of the cross section is carried out in the energy range of 10 keV–7 MeV. The black dashed curve shows the reaction rate that corresponds to the black dashed curve in Fig. 2a and takes into account only the $E1$ transition from the $S$ scattering wave.

Here, the reaction rate for results of paper [39] is shown by the blue band. In addition, the green band is presenting in Fig. 4 the results for the $\langle \sigma v \rangle$ from [38]. Both of these calculations are lower than the result obtained here by some orders of magnitude. It is seen from Fig. 4 that consideration of the first resonance essentially changes the reaction rate as against results of publications [38,39], where only the nonresonance calculations were carried out. Accounting for the two next resonances in present calculations leads only to the small increasing of the $\langle \sigma v \rangle$ at highest temperatures.

The parametrization of the reaction rate was carried out using the expression of the form [43]:

$$N_A \langle sv \rangle = \frac{a_1}{T^{2/3}} \exp\left(-\frac{a_2}{T^{1/3}}\right)\left(1 + a_3 T^{1/3} + a_4 T^{2/3} + a_5 T + a_6 T^{4/3} + a_7 T^{5/3} + a_8 T^{7/3}\right) +$$
$$+ \frac{a_9}{T^{1/2}} \exp\left(-\frac{a_{10}}{T^{1/2}}\right) + \frac{a_{11}}{T} \exp\left(-\frac{a_{12}}{T}\right) + \frac{a_{13}}{T^{3/2}} \exp\left(-\frac{a_{14}}{T^{3/2}}\right) + \frac{a_{15}}{T^2} \exp\left(-\frac{a_{16}}{T^2}\right) \qquad (9)$$

with the parameters listed in Table 6. The results of these parametrization are shown above in Fig. 4 by the red dashed curve.

Table 6. Parameters of the analytical parametrization with $\chi^2 = 0.06$ at 5% errors of the calculated total reaction rate.

| $a_1$ | $a_2$ | $a_3$ | $a_4$ | $a_5$ | $a_6$ | $a_7$ | $a_8$ |
|---|---|---|---|---|---|---|---|
| -146.8134 | 7.52023 | -1.57556E6 | 5.38408E6 | -4.16671E6 | 109516 | 564105.0 | -41092.96 |
| $a_9$ | $a_{10}$ | $a_{11}$ | $a_{12}$ | $a_{13}$ | $a_{14}$ | $a_{15}$ | $a_{16}$ |
| -417.1643 | 2.34235 | 2.75351E6 | 0.94389 | -2.82805E6 | 1.18361 | 1.26521E6 | 0.99103 |

The parametrization of the reaction rate for the nonresonance $E1$ transition shown in Fig. 4 by the black dashed curve for the $E1$ capture was done additionally. The next form was used for:

$$N_A \langle sv \rangle = \frac{a_1}{T^{2/3}} \exp\left(-\frac{a_2}{T^{1/3}}\right)\left(1 + a_3 T^{1/3} + a_4 T^{2/3} + a_5 T + a_6 T^{4/3} + a_7 T^{5/3} + a_8 T^{7/3}\right) \quad (10)$$

with parameters listed in Table 7.

Table 7. Parameters of the analytical parametrization with $\chi^2 = 0.06$ at 5% errors of the calculated reaction rate for the nonresonance $E1$ transition.

| $a_1$ | $a_2$ | $a_3$ | $a_4$ | $a_5$ | $a_6$ | $a_7$ | $a_8$ |
|---|---|---|---|---|---|---|---|
| -11.52211 | 8.41371 | -5.44959E6 | 7.78529E6 | -3.98974E6 | 201.7225 | 457623.9 | -32323.2 |

The result of this parametrization is shown in Fig. 4 by the light blue dashed curve. It is seen that, like in previous case, the description of the calculated curve by this parametrization can be considered quite reasonable.

## 8. Conclusion

It is possible to construct two-body potentials of the $p^8$Li interaction, which allow us to correctly describe the available data on characteristics of the bound state of $^9$Be in the $p^8$Li channel in the frame of the MPCM. Suggested options of the GS potentials of $^9$Be in the $p^8$Li channel allow one to obtain AC within limits of errors available for it and lead to the reasonable description of $^9$Be radii. Such potentials generally allow one to reproduce the available experimental data for total cross sections of the radiative proton capture on $^8$Li at low and ultralow energies. Obtained results for total cross sections and static characteristics of $^9$Be strongly depend of GS potential parameters of this nucleus in the $p^8$Li channel.

New results for the astrophysical $S$-factor and reaction rate have been obtained. They turned to be essentially higher than similar results of previous works [38,39], where the first resonance $1/2^-$ (0.087) was not taken into account. The parametrization of the calculated $S$-factor in the energy range to 50 keV was carried out. Further account of two other resonances $5/2^-$ (0.410) and $3/2^+$ (1.100) leads only to small changes of the reaction rate at highest temperatures. Analytical parametrization of the calculated total reaction rate was carried out. As well as the nonresonance $E1$ reaction rate was parametrized. The obtained results for the reaction rate of the proton capture on $^8$Li may lead to an essential re-estimation of the efficiencies of light nuclei in different thermonuclear processes in the Universe.

We consider the presented results as estimative, so as undoubtedly targeted measurements of the capture cross sections are needed in the range, where our model predicts the resonance behavior of cross sections, supported by the modern data on $^9$Be spectra near the proton threshold. Finally, let us note that further refinement of experimental characteristics of high-lying levels of $^9$Be also is extremely desirable.

**Acknowledgments**

This work was supported by the Ministry of Education and Science of the Republic of Kazakhstan (Grant No. AP05130104) entitled "Study of Reactions of the Radiative Capture in Stars and in the Controlled Thermonuclear Fusion" through the Fesenkov Astrophysical Institute of the National Center for Space Research and Technology of the Ministry of Digital Development, Innovation and Aerospace Industry of the Republic of Kazakhstan (RK).